\documentclass[12pt]{article}
\usepackage{makeidx}

\title{Symmetries of the dynamics and Noether theorem  in classical mechanics} 
\author{F. Strocchi \\ Theoretical Physics Group at Pisa University and INFN,\\ 56127 Pisa, Italy}

\makeglossary
\date{}
\newtheorem{Theorem}{Theorem}[section]

\newtheorem{Proposition}[Theorem]{Proposition}

\usepackage{latexsym}
\usepackage[T1]{fontenc}
\usepackage{amsmath}
\usepackage{amssymb}

\def \AO {{\cal A}({\cal O})}
\def \AO' {{\cal A}({\cal O}')}

\def \be {\begin{equation}}
\def \ee {\end{equation}}

\def \ra {\rightarrow}

\def \eqq {\equiv}

\def \d {{\delta}}
\def \eps {{\varepsilon}}

\def \F {{\cal F}}

\def \Q {{\cal Q}}

\def \d^nu {{\partial^\nu}}
\def \d^la {{\partial^\lambda}}
\def \d^o {{\partial^0}}

\def \p {{\bf p}}

\def \x {{\bf x}}

\def \Rbf {{\bf R}}

\def\doppio#1{{\rm I}\kern-.1667em{\rm #1}}

\def\Q{\text{Q}\kern-.52em
    \text{\vrule height1.5ex width.5pt depth0pt}\kern.45em}

\def\dZ{{\mathchoice {\hbox{$\Ss\textstyle Z\kern-0.4em Z$}}
{\hbox{$\Ss\textstyle Z\kern-0.4em Z$}} {\hbox{$\Ss\scriptstyle
Z\kern-0.25em Z$}} {\hbox{$\Ss\scriptscriptstyle Z\kern-0.2em
Z$}}}}

\def\dC{{\mathchoice{\hbox{$\rm\textstyle\text{\kern.35em\vrule
   height1.5ex width.5pt depth0pt\kern-.35em C}$}}
{\hbox{$\rm\textstyle\text{\kern.35em\vrule
   height1.5ex width.5pt depth0pt\kern-.35em C}$}}
{\hbox{$\rm\scriptstyle\text{\kern.35em\vrule
   height1.5ex width.3pt depth0pt\kern-.35em C}$}}
{\hbox{$\rm\scriptscriptstyle\text{\kern.35em\vrule
   height1.5ex width.2pt depth0pt\kern-.35em C}$}}}}

\makeatletter

\@addtoreset{equation}{section}

\begin{document}
\maketitle

\begin{abstract}
The aim of this note is to discuss the relation between one-parameter continuous symmetries of the dynamics, defined on physical grounds,  and conservation laws. In the Hamiltonian formulation, such symmetries of the dynamics in general leave the Hamiltonian invariant only up to a total derivative  
$ d G(q)/d t$ . In this more general case, the corresponding formulation of Noether theorem gives that the conservation law displays a sort of ano\-ma\-ly, the constant of motion being the sum of the canonical generator of the symmetry transformations plus the generator $G$ of the gauge transformation $q_i \ra q_i$, $p_i \ra  p_i - \partial G/ \partial q_i$. 

\end{abstract}

\newpage

\section{Introduction}
The deep relation between continuous symmetries and conservation laws is the basic result of Noether theorem. 

In most textbook presentations [1-3], the concept of continuous symmetry is identified with the invariance of the Lagrangian under a one-parameter continuous group of transformations of the Lagrangian variables and their time derivatives. 

In  more refined treatments [4], the invariance of the Lagrangian is required only up to a total derivative. Since the expression for the derived conserved quantity is affected by the presence of such a total derivative, the question arises about the meaning and implications of a continuous one-parameter group of symmetries for a given  physical system. 

From a physical point of view, one is led to consider as symmetries of a physical  system those transformations of the dynamical variables which leave the equations of motion invariant. 
It is part of the common wisdom that the invariance of the Lagrangian up to a total derivative implies the invariance of the equations of motion, but the converse is not emphasized in most textbooks. 

Actually, one has that {\em the invariance of the equation of motions  is equivalent  to the invariance of the Lagrangian up to a total derivative} (see e.g.[5]).
Thus,  Noether theorem for one-parameter continuous groups of transformations leaving the Lagrangian invariant up to a total derivative completely characterizes the relation between between symmetries of the dynamics and conservation laws.

The Hamiltonian counterpart of such an important relation is usually trivialized  by considering as symmetries of a physical system those transformations of the canonical variables which leave the Hamiltonian invariant. This trivially implies that the generator  of a one-parameter  continuous  group of such transformations of the canonical variables is a constant of motion (since it has vanishing Poisson brackets with the Hamiltonian).   

Actually, a more refined analysis is required. As discussed above, the invariance of the  equations of motion implies the invariance of the Lagrangian up to a total derivative and therefore one has to discuss the Hamiltonian counterpart of this relation. This is the object of the present note.\goodbreak

The result is that a one-parameter continuous group of (possibly time dependent) transformations of the dynamical variables which leave  their time evolution invariant and therefore leave the Lagrangian invariant up to a total derivative, $ d G(q)/ d t$, induces the following infinitesimal transformations of the Hamiltonian 
\be{ \delta H \eqq H(q', p', t) - H(q, p, t) =  \eps \frac{\partial F}{\partial t} + \eps 
\frac{d\, G(q)}{d t},}\ee  
where $F$ is the (possibly time dependent) generator of the corresponding canonical transformations $q, \,p$ $\ra q', \,p'$.

In this more general case, the Hamiltonian formulation of  Noether theorem  gives the following conservation law
\be{  \frac{d {\cal Q}}{ d t} \eqq  \frac{d}{ d t} (F + G) = 0.}\ee
Hence, in order to get  a conserved quantity one has to add the function $G$ to the generator $F$ of the canonical transformations. 

Since the addition of a total derivative to the Lagrangian does not change the dynamics of the Lagrangian variables $q, \dot{q}$, it leaves invariant all the observables $F(q, \dot{q})$ and has therefore the meaning of a {\em gauge transformation} (this point of view is shared by Ref.\,[4], pp.124-127,  however, with different  Hamiltonian version of Noether theorem). 

In terms of the canonical variables the addition of the total derivative implies the following transformation of the canonical variables
\be{q_i \ra q_i, \,\,\,\,\,p_i \ra p_i - \frac{ \partial  G}{ \partial q_i},}\ee    
which  changes the relation between the conjugate momentum $\p_i$ and the time derivative 
$\dot{q}_i$ of the position. Equation (1.3) states that $G$ is the canonical generator of such a gauge transformation. It is worthwhile to recall that for a particle in a magnetic field $\x$ and $\dot{\x}$ are observable (gauge invariant) quantities, but $p_i \eqq \dot{x}_i + (e/c) A_i$ is not.

In conclusion for one-parameter continuous groups of transformations, which leave the dynamics invariant, but leave the Lagrangian or the Hamiltonian  invariant only up to a total derivative, the conservation laws displays a sort of {\em anomaly}, the conserved quantity being the sum of the generator of the corresponding canonical transformation plus the generator $G$ of the gauge transformation (1.3).

\section{Symmetries of the dynamics and transformation of the Hamiltonian}
  By {\em symmetry of the dynamics} we mean a transformation of the dynamical variables such that their  equations of motion are invariant. In the Lagrangian formulation, the dynamical variables are the Lagrangian coordinates $q_i$ and their time derivatives $\dot{q}_i$ and a transformation
\be{ q_i \ra q'_i(q, t), \,\,\,\,\,\,\dot{q}_i  \ra \dot{q}'_i(q, \dot{q}, t), }\ee
is a symmetry of the dynamics if it leaves the equations of motion $\ddot{q}_i = F_i(q, \dot{q}, t)$ invariant, i.e. 
\be {\ddot{q}\,' = F_i(q', \dot{q}', t).}\ee
Then, one has a complete characterization of the symmetries of the dynamics in terms of invariance properties of the Lagrangian [5]:
\begin{Proposition}  The invariance of the Lagrange equations under a (possibly  time dependent) transformation
of the Lagrangian variables $q_i \ra q_i', $  $\dot{q}_i \ra \dot{q}'_i$ is equivalent to the invariance of the Lagrangian up to a total derivative
\be{ L'(q', \dot{q}', t ) = L(q', \dot{q}', t) - \frac{d G(q)}{ d t}.}\ee  
\end{Proposition}
\vspace{1mm}

Since the Lagrangian transforms covariantly under a change of the Lagrangian coordinates,
namely 
\be{ L'(q', \dot{q}', t) = L(q, \dot{q}), t) ,}\ee
eq.\,(2.3) (and therefore the symmetry of the dynamics) is equivalent to 
\be{ L(q', \dot{q}', t ) = L(q, \dot{q}, t) + \frac{d G(q)}{ d t}.}\ee
 The next step is to characterize the transformation properties of the Hamiltonian
under a transformation of the Lagrangian variables which leave the Lagrangian invariant up to a total derivative.

\begin{Proposition} A transformation of the Lagrangian variables 
\be{ q_i \ra q_i'(q, t), \,\,\,\,\,\dot{q}_i \ra \dot{q}'_i(q, \dot{q}, t), }\ee
such that  the Lagrangian is invariant up to a total derivative, eq.\,(2.3), defines a canonical transformation of the canonical variables
\be{ q_i \ra q'_i, \,\,\,\,\,\,\,\,p_i \ra p'_i,}\ee
such  that  
\be{ H'(q', p', t) = H(q', p', t) - \frac{d G}{d t}.}\ee 
\end{Proposition}
{\bf Proof}. In fact, one has (sum over repeated indices being understood)
$$ H(q', p',t )= \dot{q}'_i\, p'_i - L(q', \dot{q}', t) = \dot{q}'_i\, p'_i - L(q, \dot{q}, t) - \frac{d G(q)}{ d t} = $$ 
$$ = \dot{q}'_i p'_i - L'(q', \dot{q}', t) -\frac{d G(q)}{ d t} = H'(q', p', t) 
+ \frac{d G(q)}{ d t},$$
where we have used eq.\,(2.5) and eq.\,(2.4).

For a one-parameter continuous groups of canonical transformations, the infinitesimal variations of the canonical variables $q, p$ are of the form
\be{ \delta q_i = \eps \{ q_i, \,F \} = \eps \frac{\partial F(q, p, t)}{ \partial p_i}, \,\,\,\,\,\,\,\delta p_i  =  \eps \{ p_i, \,F \} = - \eps \frac{\partial F(q, p, t)}{ \partial q_i},}\ee
where $F(q, p, t)$ is the generator of the canonical transformation and $\{\,\,,\,\,\}$ denotes the Poisson bracket. 

Clearly, a one-parameter group of  symmetries of the dynamics is non-trivial, provided 
$\delta q_i \neq 0$.  
Then , one has:
\begin{Theorem} {\bf Noether theorem. Hamiltonian form}.\index{Noether theorem!Hamiltonian form}
 To each one-parameter 
group  of (non-trivial)  symmetries of the dynamics,  so that  in the Hamiltonian formulation the Hamiltonian is invariant up to a total derivative, eq.\,(2.8), there corresponds the following constant of motion 
\be{{\cal Q} \eqq F + G,}\ee
$F$ being the canonical generator of the symmetry transformations, eqs.\,(2.9).
\end{Theorem}
{\bf Proof}. The first step is to derive the Hamiltonian analog of eq.\,(2.5), i.e.
 one must relate 
$H'(q', p', t)$ to $H(q, p, t)$. 

Contrary to the Lagrangian case, for time dependent transformations $H'(q', p', t) \neq H(q, p, t)$; actually one has (see e.g. [5])
\be{H'(q', p', t) =  H(q, p, t) + \frac{\partial \F}{ \partial t},}\ee
where $\F$ is a generating function of the canonical transformation (2.9).

Then, the  expansion of $\F$ to first order in $\eps$, in eq.\,(2.11) gives 
\be{H'(q', p', t) =  H(q, p, t) + \eps \frac{\partial F}{ \partial t},}\ee
and eq.\,(2.8) is equivalent to 
\be{ \delta H \eqq H(q', p', t) - H(q, p, t) = \eps \frac{\partial F}{\partial t} + \eps \frac{\partial G}{ \partial t}.}\ee   
Now, on one side,  one has
$$  \delta H = \eps \,\left( \frac{\partial H}{\partial q_i} 
\frac{\partial F}{\partial p_i} -\frac{\partial H}{\partial p_i} 
\frac{\partial F}{\partial q_i}\right)  = 
\eps\, \{ H, \,F \} = - \eps \left( \frac{ d F}{ d t}-  \frac{\partial F}{ \partial t} \right), $$
and, on the other side, by eq.\,(2.13), one has
$$ \delta H =  \eps \frac{\partial F}{\partial t} 
+ \eps \frac{d\, G(q)}{d t}.$$
Hence, it follows that
$$-\,\frac{d F}{ d t} =  \frac{d G(q)}{ d t}.$$
i.e. ${\cal Q} = F + G $ is a constant of motion. 

\vspace{2mm} 
The above Theorem allows to derive the constant of motion associated to a one-parameter group of symmetries of the dynamics, without recourse to the Lagrangian formulation;
one has to check only the transformation properties of the Hamiltonian, leaving open its invariance up to a total derivative, eq.\,(2.8).

The conclusion is that canonical generator $F$ of a symmetry of the dynamics need not be a constant of motion. The point is that the invariance of the dynamics requires the invariance of the Hamiltonian only up to a total derivative $d G/ d t$, with $G$ the generator of a  gauge transformation, and ,in general, only the sum $F + G$ is a constant of motion (anomaly). 

On the other hand,  the standard treatment of the Hamiltonian version of Noether theorem,  identifies the symmetries as those which leave the Hamiltonian invariant, implicitly giving a backbone role to  the  relation between the canonical momentum $p_i$ and the ''velocity'' $\dot{q}_i$. 
Since such a relation is not invariant under gauge transformations,
it  should not be allow to affect the characterization of the symmetries of the system, which rather have a physical (observable) content. 

In fact, the aim of the above Theorem is to relate the physical property of invariance of the dynamics to the existence of a physical constant of motion.

\vspace{15mm} 
REFERENCES
\vspace{2mm}

\noindent 
[1]H. Goldstein, C. Poole and J. Safko, {\em Classical Mechanics}, 3rd. ed., Pearson 2006

\vspace{1mm}
\noindent
[2] V.I. Arnold, {\em Mathematical Methods of Classical Mechanics}, 2nd. ed. Springer 1989

\vspace{1mm}
\noindent 
[3] A. Fasano and S. Marmi, {\em Analytical Mechanics. An Introduction}, Oxford Univ. Press 2006  

\vspace{1mm}
\noindent
[4] J.V. Jos\'{e} and E. Saletan, {\em Classical Dynamics. A Contemporary Approach}, Cambridge Univ. Press 2006  

\vspace{1mm}
\noindent 
[5] \,F. Strocchi, {\em A primer of Analytical Mechanics}, Springer UNITEXT

\end{document}